\PassOptionsToPackage{table,xcdraw}{xcolor}
\documentclass[sigconf]{acmart}
\usepackage[noend]{algpseudocode}
\usepackage{graphicx}
\usepackage{algorithmicx,algorithm}
\usepackage{url}
\usepackage{todonotes}
\usepackage{hyperref}
\usepackage{multirow}
\usepackage{makecell}
\usepackage[table,xcdraw]{xcolor}
\usepackage{pdfpages}
\usepackage{float}
\usepackage{placeins}

\usepackage{xspace}

\definecolor{lightcgray}{rgb}{0.404, 0.992, 0.604}
\definecolor{lightred}{rgb}{1, 0.8, 0.788}

\AtBeginDocument{%
  \providecommand\BibTeX{{%
    \normalfont B\kern-0.5em{\scshape i\kern-0.25em b}\kern-0.8em\TeX}}}

\setcopyright{acmcopyright}

\copyrightyear{2023}
\acmYear{2023}
\setcopyright{rightsretained}
\acmBooktitle{Proceedings}\acmDOI{xx.xxx/xxx.xxx}
\acmISBN{xxxxx}

\begin{document}

\title{TutorLoop: Regulating Student Learning Behaviors via Sensor-in-the-Loop Generative Feedback}

\author{Songlin Xu}
\email{songlin.xu@polyu.edu.hk}
\affiliation{%
  \institution{The Hong Kong Polytechnic University}
  \city{Hong Kong}
  \country{China}
}

\author{Xinyu Zhang}
\email{xyzhang@ucsd.edu}
\affiliation{%
  \institution{University of California, San Diego}
  \city{San Diego}
  \state{California}
  \country{USA}
}

\renewcommand{\shortauthors}{Xu, et al.}

\renewcommand{\shortauthors}{xxxx, et al.}

\begin{abstract}

We present TutorLoop, a sensor-in-the-loop system that regulates student learning behaviors by delivering adaptive feedback based on real-time cognitive states. Unlike prior large language model (LLM) tutors that directly depend on scenario-specific content, TutorLoop operates on sensor-derived signals captured via webcams. Moreover, unlike direct cognitive-to-feedback mappings that are short-sighted, the system employs a deep reinforcement learning (DRL) agent to optimize the feedback type across the entire learning process. Finally, another LLM tutor refines feedback into human-like, context-aware messages. We evaluate TutorLoop in a large-scale user study (N=187), where a model trained offline is directly applied to a new learning task without retraining. Results show that TutorLoop provides less frequent yet more effective interventions, improving attention, reducing workload, increasing engagement, and ultimately enhancing learning outcomes. These findings highlight the potential of closed-loop, sensor-driven feedback for scalable human-AI integrated systems to support learning.

\end{abstract}



\begin{teaserfigure}
  \includegraphics[width=\textwidth]{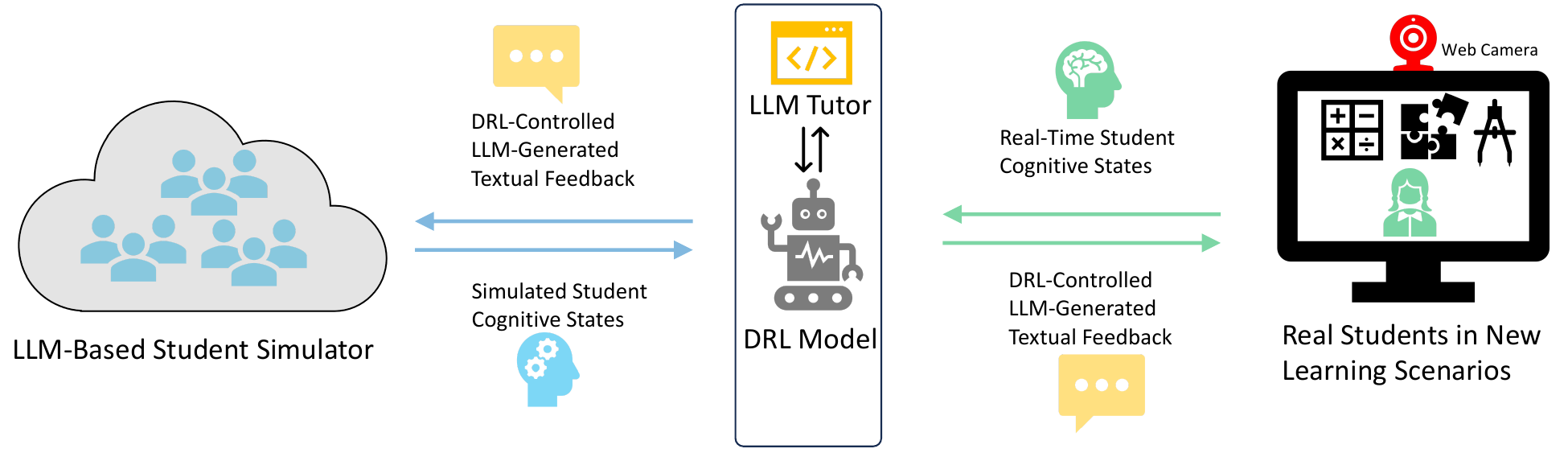}
  \caption{Our framework. TutorLoop leverages students' real-time cognitive states, captured via webcams, to regulate learning behaviors and enhance outcomes. A DRL model controls an LLM tutor to generate adaptive feedback. The DRL model is first trained with the student simulator  and then directly applied to real students in new learning tasks without model updates.
}
  \Description{
  A schematic diagram shows the TutorGen framework in two parallel paths: on the left, simulated student cognitive states flow from the student simulator into a DRL model, which then controls LLM-generated textual feedback; on the right, real students’ cognitive states captured by webcams are processed in real time by the same DRL-controlled LLM tutor to provide feedback. The diagram emphasizes that the DRL model is first trained virtually with the student simulator and then directly applied to real students without further updates.
  }
  \label{fig: teaser}
\end{teaserfigure}

\maketitle

\section{Introduction}
\label{sec:intro}



Online education plays a crucial role not only as a strategic response to disruptions such as natural disasters and public health emergencies \cite{uscher2018school}, but also as a universally accessible platform  \cite{mcloughlin2001inclusivity} for students who face barriers to attending traditional in-person classes. However, online education suffers from intrinsic limitations. In particular, online platforms often lack mechanisms to sustain student engagement, as learners naturally lose attention in the absence of in-person instructors \cite{peper2021avoid}.  
Recent advances in large language models (LLMs) show promise in addressing this gap. LLMs have been applied to train tutors or serve directly as tutors. It can deliver textual feedback to enhance engagement \cite{pan2025tutorup,bodonhelyi2025passive}, augment problem solving \cite{zha2025mentigo}, generate questions \cite{kang2025tutorcraftease,lu2024generative}, and support personalized learning \cite{pal2024autotutor,phung2024automating,jin2025teachtune,hou2024codetailor}. However, most prior work grounds AI tutors in specific courses or task contexts \cite{xu2023augmenting}, which limits their ability to transfer across new learning scenarios.

We propose a general \textit{sensor-in-the-loop} system that leverages student cognitive states (attention and workload), captured through ubiquitous sensors (e.g., web cameras), as the observation space for generating LLM-based feedback. Unlike prior approaches, our system does not require course content as input, allowing it to generalize across diverse learning settings and task domains.  
However, directly mapping cognitive states to feedback tends to be short-sighted. 
For instance, intervening whenever attention drops may inadvertently increase workload and ultimately impair performance, based on the Yerkes–Dodson law \cite{yerkes1908relation}. An effective strategy must therefore balance attention and workload while considering long-term learning outcomes, which are shaped by the cumulative learning process rather than isolated moments.

In this paper, we introduce \textbf{TutorLoop}, a tutoring framework in which a deep reinforcement learning (DRL) model controls the LLM tutor to provide far-sighted feedback. TutorLoop optimizes cumulative learning rewards by modeling cognitive states over the entire learning process. However, two key challenges remain.

The first major challenge is how to generate persuasive and effective feedback that students are willing to follow. The design space of LLM-generated feedback is vast. It could involve open-ended reflective prompts that encourage self-exploration \cite{zheng2024charting}, concrete suggestions that direct attention to specific course content \cite{xu2025peeredu}, and emotionally framed cues that influence motivation or confidence \cite{leong2024putting}, etc. Each feedback form can further vary in structure, intensity, and timing.
Directly fine-tuning LLMs to explore this space is computationally expensive and requires massive interaction data with real students, making it impractical.  

To address this, we narrow the feedback space by grounding it in cognitive science principles. The Yerkes–Dodson law \cite{yerkes1908relation} posits that performance improves with arousal up to an optimal point but deteriorates when excessive workload induces over-arousal. We use attention signals captured by sensing data as a proxy for arousal. Following this law, our design includes two core feedback types: (a) \textit{attentive feedback}, which reminds students to increase attention when it drops, and (b) \textit{relief feedback}, which suggests relaxation when workload is high. In addition, inspired by findings that praise reinforces engagement \cite{partin2009using}, we add (c) \textit{encouragement feedback} to acknowledge desirable behaviors such as sustained attention. Combined with \textit{no feedback} when intervention is unnecessary, these yield four discrete feedback types.

We employ a DRL model to adaptively select which type of feedback to deliver based on real-time cognitive states and the long-term trajectory of the learning process.  We further employ an LLM tutor to refine how the selected action is conveyed (w.r.t. wording, tone, contextualization, etc.), 
so as to maximize persuasiveness and responsiveness. 
The LLM tutor is designed to counteract the habituation effect \cite{grissom2009habituation}, i.e., reduction of attention and responsiveness against repeated exposure to the same type of feedback stimuli. 
In addition, to enhance student compliance, the LLM tutor is tailored to generate contextualized feedback based on a student's past behaviors \cite{sugai2012contextual}, as students who perceive that their actions are being monitored are more likely to follow the feedback \cite{butler1995feedback}. 

The second major challenge lies in training the DRL model. Ideally, the training would require massive amounts of interactive online training data, which is difficult to collect in real time with human learners. 
We overcome this challenge through an LLM-based student simulator using an architecture similar with EduAgent \cite{xu2024eduagent}, which generates cognitive states in response to course materials, based on few-shot demonstration from public datasets. 
With this simulator, we train the TutorLoop model on public datasets in virtual environments and then directly apply it \textbf{without retraining} to a new self-learning task in a large-scale study ($N=187$). This demonstrates the transferability of our sensor-integrated system. Specifically, while simulators may draw on course materials to generate contextual behaviors, the DRL model relies solely on cognitive states as input, independent of course materials.

Results show that TutorLoop significantly improves attention, reduces workload, increases active engagement (e.g., slide switching), and ultimately enhances learning outcomes compared to three baseline groups.  
Interestingly, our study reveals that while attention alone is weakly correlated with learning outcomes, workload is strongly negatively correlated. This suggests that feedback strategies narrowly focused on increasing visual attention may not substantially improve learning and may even introduce cognitive overload. Instead, feedback should account for the tradeoff between attention and workload, consistent with the Yerkes-Dodson law \cite{yerkes1908relation}. Furthermore, our study shows that directly using LLMs to generate feedback based on each updated cognitive state yields only limited improvements in learning outcomes compared to our approach. A likely reason is that although LLMs can condition on the entire history of cognitive states, they are not explicitly optimized with a reward function as in deep reinforcement learning. As a result, the generated feedback may be suboptimal. Finally, our TutorLoop system achieves superior outcomes with fewer feedback interventions, underscoring the effectiveness of adaptive, judiciously timed feedback over frequent, intrusive prompts.

In summary, our key contributions are as follows:
\begin{itemize}
    \item We propose a general sensor-in-the-loop framework where a DRL model controls an LLM tutor to regulate students' online cognitive states and augment learning performance. By relying solely on sensor-derived information rather than course-specific materials, TutorLoop can be directly applied in a new task after pre-training in a public dataset.  
    \item We show that student simulators can be directly used to train AI models to address the data scarcity bottleneck for human-in-the-loop training.
    \item Our between-subject experiment with N=187 participants not only demonstrates the unique advantage of TutorLoop in regulating student behaviors and augmenting learning outcomes, but also reveals insights about limitations in directly applying LLMs in regulating students' behaviors.  
\end{itemize}

\section{Related Work}
\label{sec: related work}

Our work mainly draws inspirations from and advances the knowledge in the following three categories of research. 

\subsection{LLMs in Education}

LLMs have been widely explored in education, serving as tutors or support tools to enhance engagement (TutorUp \cite{pan2025tutorup}, \cite{bodonhelyi2025passive}), augment problem solving (Mentigo \cite{zha2025mentigo}), generate practice questions (TutorCraftEase \cite{kang2025tutorcraftease}), and support personalized learning \cite{pal2024autotutor,phung2024automating,hou2024codetailor,shao2025unlocking,zhu2025exploring}. They have also been used with simulated students to evaluate tutors or generate feedback (TeachTune \cite{jin2025teachtune}, Generative Students \cite{lu2024generative}, \cite{nair2024closing}). Beyond tutoring, LLM-powered agents \cite{prasad2025exploring} have supported tasks (Mathemyths \cite{zhang2024mathemyths}, DevCoach \cite{wang2024devcoach}, \cite{jin2024teach}, \cite{liang2023let}) such as recommending concepts \cite{li2024learning}, providing instructions \cite{vadaparty2024cs1},  delivering feedback \cite{matelsky2023large}, or assisting instructors in content design (ReadingQizMaker \cite{lu2023readingquizmaker}).

\subsection{LLMs for Student Simulation}

Researchers have explored various ways of leveraging LLMs for student simulation. Prior work has largely focused on knowledge tracing to predict learning skills from historical data \cite{fu2024sinktstructureawareinductiveknowledge,jung2024clst,li2024integrating,lee2024monacobert,lee2024language,zhang2024predicting,yu2024eckt}, or on generative agents that simulate classroom interactions \cite{yue2024mathvcllmsimulatedmulticharactervirtual,zhang2024simulating,chen2023agentverse,jinxin2023cgmi,mannekote2024can}. These efforts have supported adaptive practice generation \cite{cui2023adaptive}, interactive TA training \cite{markel2023gpteach}, and personalized tutoring \cite{sarshartehrani2024enhancing}, and have also been extended to domains such as self-reflection \cite{kumar2024supporting}, and growth mindset development \cite{kim2024exploring}. 

Recent work has moved toward more contextual behavioral simulation in online education. For example, Classroom Simulacra \cite{xu2025classroom} builds contextual student agents, while EduAgent \cite{xu2024eduagent} is most closely related to our work in simulating cognitive states. Since student simulation is not the core contribution of this work, we adopted the similar architecture as EduAgent to build the student simulator, serving as the virtual environment to train our deep reinforcement learning model in the TutorLoop framework.

\subsection{LLMs to Enhance Student Learning}

Recent HCI research has explored a wide range of LLM-powered systems to support and enhance student learning. Several works focus on domain-specific applications, where LLMs generate explanations, exercises, or feedback tailored to particular subjects. For instance, CodeTailor provides personalized Parsons puzzles to support programming engagement \cite{hou2024codetailor}, Rubikon leverages AR and LLMs for Rubik’s Cube tutoring \cite{ren2025rubikon}, and eXplainMR generates real-time textual and visual explanations to facilitate ultrasound training \cite{wang2025explainmr}. Similarly, Surgment supports surgery learning through segmentation-enabled semantic feedback \cite{wang2024surgment}, and explainable coding systems such as PaTAT enable interactive rule synthesis in programming education \cite{gebreegziabher2023patat}.  

Other research investigates more general applications of LLM-generated feedback. Lu et al. studied how teaching assistants evaluate LLM feedback for economics essays \cite{lu2025exploring}, while Doherty et al. designed an LLM-based educational jigsaw agent to support teamwork learning \cite{doherty2025piecing}. Co-design approaches have also been employed to explore multimodal generative AI in educational settings \cite{prasad2025exploring}.  
Most relevant to our work are systems that directly investigate adaptive feedback for regulating student behaviors. Shochcho et al. examined how LLM-based tutors can improve engagement and outcomes in Python learning \cite{shochcho2025improving}, and Bassen et al. applied reinforcement learning to optimize the scheduling of educational activities \cite{bassen2020reinforcement}. TutorUp \cite{pan2025tutorup} further explored the use of simulated students to train tutors for addressing engagement challenges in online learning.  

Despite these advances, most prior work grounds AI tutors in specific courses or task contexts \cite{xu2023augmenting}, limiting their transferability across diverse learning scenarios, since reliance on course materials does not guarantee successful application to new courses or tasks. In contrast, we propose a general sensor-in-the-loop framework that generates adaptive feedback solely from cognitive states, without course-specific inputs, thereby enabling generalization across tasks and domains. Our user study corroborates this advantage, demonstrating that a model trained on simulated students from public datasets can be directly applied to support learning for new real students in a novel task context.

\begin{figure*}
\centering
\includegraphics[width=0.96\linewidth]{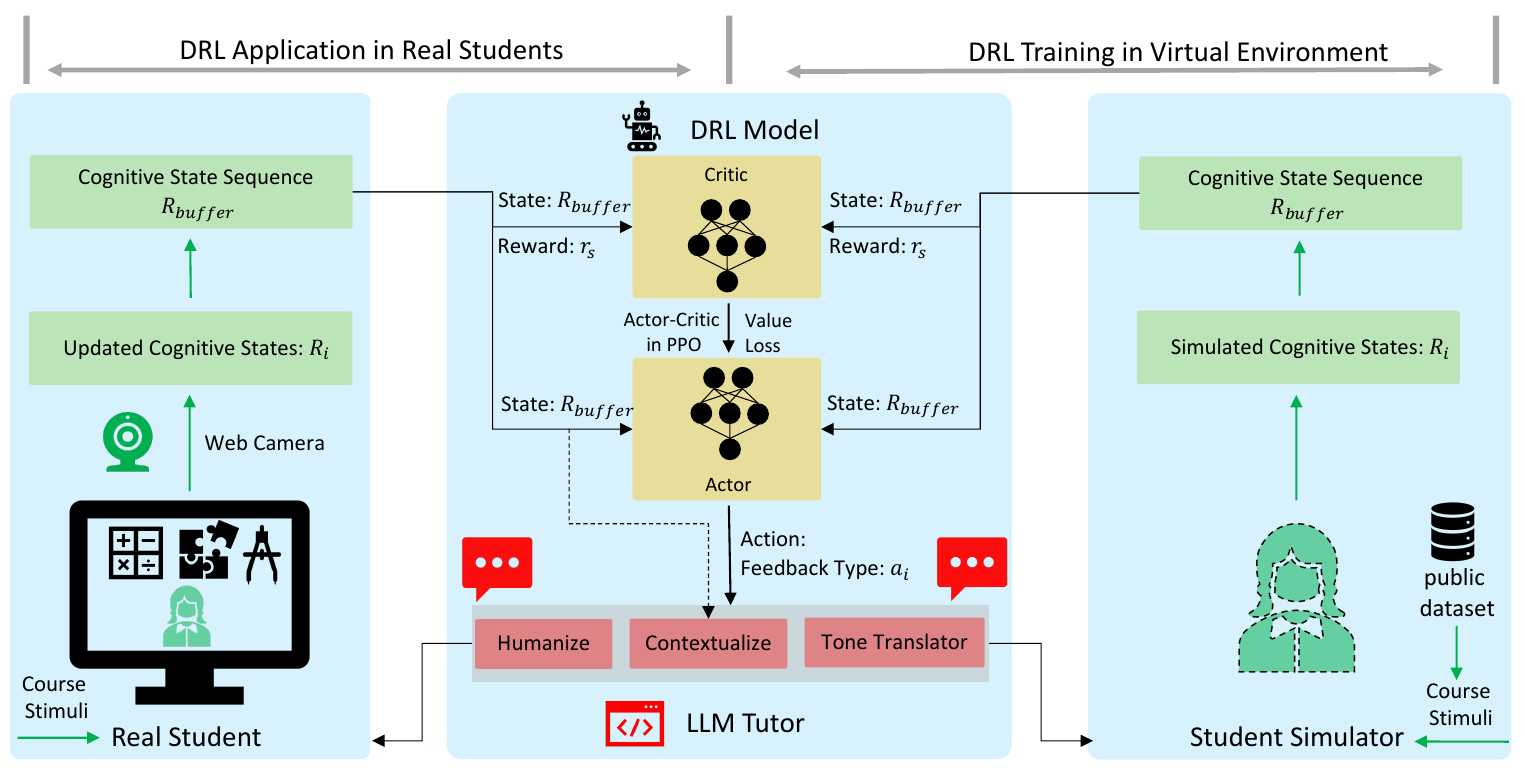}
\caption{
The TutorLoop framework details. Note that course stimuli are only used to stimulate real/simulated students to generate cognitive states, but the DRL model does not need course stimuli as input.
}
\Description{
A block diagram details the full framework flow. On the left, public datasets provide input to the student simulator, which simulates cognitive states. These states feed into a DRL training pipeline (actor-critic PPO model) with buffer states, actions (feedback types), and rewards. The pipeline shows how contextualization, tone translation, and humanization modules in the LLM tutor refine feedback. On the right, arrows extend to real student application with webcam-based sensing. 
}
\label{fig: framework detail}
\end{figure*}

\section{TutorLoop: System Design}
\label{sec: TutorLoop}

As illustrated in Fig.~\ref{fig: teaser}, the system comprises two components: \textit{TutorLoop} and \textit{the student simulator}. TutorLoop is a closed-loop feedback framework that receives students' cognitive states (attention and workload, measured via webcam) and generates textual feedback. The feedback is controlled by a deep reinforcement learning (DRL) model and refined by a large language model (LLM) tutor, with the objective of regulating learning behaviors and improving outcomes. The student simulator serves as a digital twin, simulating student behaviors in response to course materials and instructional feedback. TutorLoop is first trained in this virtual environment through interaction with the student simulator, after which the trained DRL agent is deployed to engage with real students on novel tasks without relying on course materials or further model adaptation.

\subsection{Multi-Modal Sensing}
We employ web cameras as sensors to capture students' cognitive states, as they are widely accessible, embedded in personal computers, and thus suitable for scalable deployment. Specifically, we adopt the open-source JavaScript library \textit{WebGazer} \cite{papoutsaki2016webgazer} for eye tracking and face detection. WebGazer operates directly on the client side, thereby preserving user privacy. The collected gaze data are represented as $(x, y, w, h, t)$, where $x$ and $y$ denote the screen coordinates of each gaze point, $w$ and $h$ denote screen width and height, and $t$ is the timestamp. Gaze points are normalized according to the user's screen size.  

As noted in Section~\ref{sec:intro}, our framework accounts for the trade-off between attention and workload, grounded in the Yerkes–Dodson law \cite{yerkes1908relation}. To estimate both measures, we partition gaze data into one-second interval and compute attention and workload within each window. These cognitive states are updated on a per-second basis.

\textbf{Attention.}
Attention captures the extent to which students focus on course materials during learning. Following prior work \cite{xu2025peeredu}, we compute the percentage of gaze points falling within the primary content area (e.g., slides or videos) rather than peripheral or blank regions of the screen, within each one-second interval.

\textbf{Workload.}
Workload reflects students' cognitive load during learning. Consistent with existing practice \cite{xu2025peeredu}, we measure workload using gaze entropy within each one-second interval, as the dispersion of gaze points has been shown to correlate with cognitive load during task performance \cite{di2016gaze}.

\textbf{Active Engagement.}
In our learning task scenario (Section~\ref{subsec:task}), students advance through course materials by clicking buttons to switch slides. We therefore capture this modality as an indicator of active engagement, quantified by the frequency of slide switching during self-learning.  

While attention and workload serve as the primary cognitive state inputs for our feedback generation model, active engagement is only recorded for later evaluation. We exclude it from model input, as not all learning tasks provide comparable data like button clicking behaviors, which would limit the system generalizability.

\subsection{Feedback Design}
\label{subsec: feedback design}

Designing feedback that students not only receive but also act upon remains a non-trivial challenge. While LLMs can in principle generate diverse forms of feedback, including reflective prompts that foster self-exploration \cite{zheng2024charting}, concrete task-oriented suggestions \cite{xu2025peeredu}, and emotionally expressive cues that influence motivation \cite{leong2024putting}, the breadth of this design space makes it difficult to determine which types meaningfully improve learning. Feedback may vary in content, tone, and timing, yet the relationships between these variations and learning outcomes remain poorly understood. Exhaustively fine-tuning LLMs to explore this space would be prohibitively expensive and demand extensive real-world student data.  

To address this challenge, we draw on the Yerkes-Dodson law \cite{yerkes1908relation}, which posits that performance improves with arousal only up to an optimal point, after which overload diminishes effectiveness. We use attention signals captured by sensing data as a proxy for arousal. 
Following this law, our design includes two core feedback types: (a) \textit{attentive feedback}, which reminds students to increase attention when it drops, and (b) \textit{relief feedback}, which suggests relaxation when workload is high. In addition, inspired by findings that explicit recognition and praise can reinforce engagement \cite{partin2009using}, we add (c) \textit{encouragement feedback} to acknowledge desirable behaviors such as sustained attention. Combined with \textit{no feedback} when intervention is unnecessary, these yield four discrete feedback types. By grounding feedback types in measurable cognitive states rather than course content, this design facilitates generalization across learning environments and task domains.  

More specifically, we initiate the feedback types based on the following prompts: (a) \textit{attentive feedback}: ``Try to be more attentive in learning!''; (b) \textit{relief feedback}: ``It seems that you are kind of anxious. Try to take a deep breath to relax yourself!''; and (c) \textit{encouragement feedback}: ``Your engagement is great! Keep your engagement!'' Once the DRL model selects a feedback type, the chosen message is refined by the LLM tutor before being delivered to the student.

\begin{figure}
\centering
\includegraphics[width=1\linewidth]{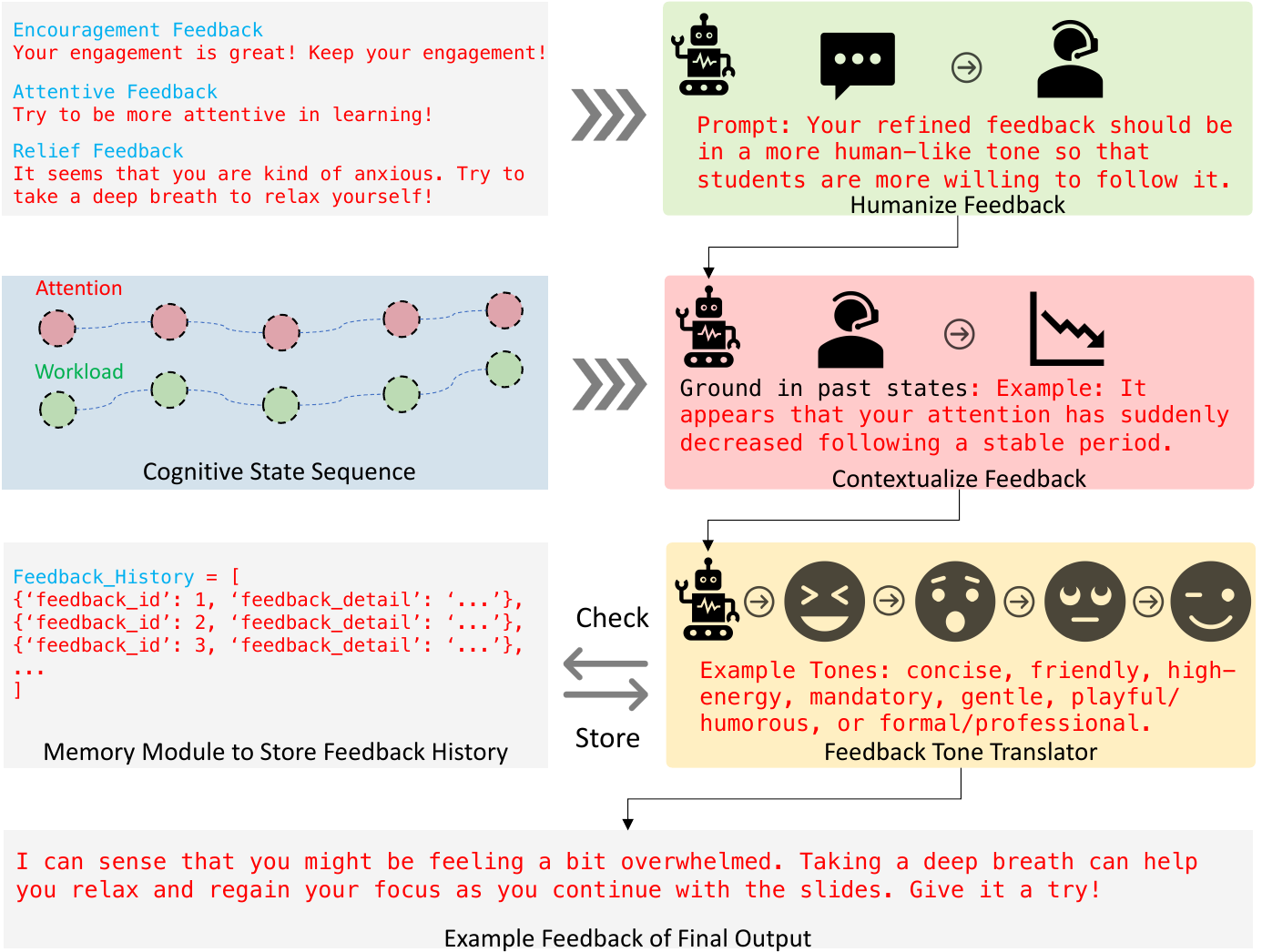}
\caption{
Details of our LLM tutor to refine the feedback, selected by the DRL model.
}
\Description{
A layered diagram illustrates the LLM tutor refinement process. At the top, three core actions (attentive, relief, encouragement feedback) are shown with example messages such as “Try to be more attentive in learning!” and “Your engagement is great! Keep your engagement!” These are stored in a feedback history memory. Refinement modules labeled “Humanize Feedback,” “Contextualize Feedback,” and “Feedback Tone Translator” process the feedback, producing varied tones such as concise, friendly, or playful. An example refined relief feedback reads: “I can sense that you might be feeling a bit overwhelmed. Taking a deep breath can help you relax and regain your focus as you continue with the slides. Give it a try!”
}
\label{fig: tutor refine detail}
\end{figure}

\subsection{Deep Reinforcement Learning (DRL) Model}

\subsubsection{Action and Observation Spaces}
As described in Section~\ref{sec:intro}, the DRL model adaptively selects feedback types (Section~\ref{subsec: feedback design}) to regulate student learning behaviors based on real-time cognitive states. Real or simulated students constitute the environment with which the DRL agent interacts. Each interaction occurs within a fixed time interval $T$, during which the most recent cognitive states form a sequence of five units that define the observation space. Each unit comprises a pair of normalized values for attention and workload (ranging from 0 to 1). Based on this sequence, the DRL agent selects one of the four feedback types in order to maximize attention while minimizing workload across the learning process. The action space is therefore defined as a discrete set of four feedback types.

\subsubsection{Terminal State}
A terminal state is reached when the DRL agent completes the maximum number of steps in an episode. To mirror real user studies, we set this maximum step number $N$ to correspond to the study duration. Upon reaching a terminal state, a new episode begins, analogous to initiating a new study session with another participant. Each episode can thus be interpreted as representing one participant's study session.

\subsubsection{Reward Function}
To account for individual differences in learning behaviors \cite{schmeck1988individual}, the reward at each step $r_s$ is defined in terms of relative changes in attention ($\Delta_{Ra}$) and workload ($\Delta_{Rw}$), compared against a student's initial baseline values ($\hat{Ra}$ and $\hat{Rw}$). The reward encourages the agent to increase attention while reducing workload, as expressed in Equation~\ref{eq:reward}:

\begin{equation}
    r_s = \Delta_{Ra}/\hat{Ra} - \Delta_{Rw}/\hat{Rw}
    \label{eq:reward}
\end{equation}

\subsubsection{Model Training}
We employ Proximal Policy Optimization (PPO) \cite{Schulman2017ProximalPO} with a multilayer perceptron (MLP) policy for model training. The DRL model is implemented using PyTorch \cite{NEURIPS2019_9015}, Stable Baselines3 \cite{stable-baselines3}, and Gym \cite{brockman2016openai}.  

As outlined in Section~\ref{sec:intro}, training is conducted through interactions with the student simulator, powered by the EduAgent dataset \cite{xu2024eduagent}, until convergence at approximately 10,000 steps. 
Notably, the DRL model is trained only with simulated students, whose cognitive states are simulated from example behaviors and course materials in the EduAgent dataset. When applied to new real students in novel tasks, the DRL model is not retrained. The rationale is that the model learns from general sensory behavioral patterns such as cognitive state changes without relying on specific course materials. Because simulators can generate large amounts of virtual data comprising a wide and diverse set of cognitive trajectories, this training corpus captures most possible patterns of cognitive state change, including those likely to occur in future students. As a result, the DRL model acquires prior knowledge that enables it to handle cognitive state dynamics even for unseen learners.
A user study demonstrating the effectiveness of the DRL model, along with detailed descriptions of baseline groups, is presented in Section~\ref{sec: TutorLoop Evaluation}.

\subsubsection{Student Simulation Environment}

As depicted...

\subsection{LLM Tutor for Feedback Refinement}

As discussed in Section~\ref{sec:intro}, delivering predefined feedback alone may lead to diminished effectiveness due to the habituation effect \cite{grissom2009habituation}, whereby repeated exposure to identical messages reduces responsiveness. The role of the LLM tutor is therefore not to select the feedback type, but rather to refine how the action prescribed by the DRL model is communicated to students, thereby improving responsiveness and adherence. For example, if the DRL agent prescribes relief feedback, the LLM tutor generates varied phrasings of the same message while preserving the core action (e.g., ``relax'') to maximize persuasiveness.  

The design of the LLM tutor is grounded in three strands of prior evidence. First, the dishabituation effect \cite{steiner2014mechanism} suggests that introducing variation, such as changes in tone, can restore attention and responsiveness. Second, contextualizing feedback based on a student's prior cognitive states increases compliance \cite{sugai2012contextual}, as students are more likely to follow feedback when they perceive that their behaviors are being monitored \cite{butler1995feedback}. Third, humanizing feedback, rather than presenting it in a purely directive style, has been shown to increase willingness to comply \cite{pacansky2020humanizing}.  

Building on these findings, the LLM tutor implements a multi-step refinement framework to humanize, contextualize, and diversify feedback tone. As illustrated in Fig.~\ref{fig: tutor refine detail}, LLMs are first prompted to humanize the initial version of feedback from Section~\ref{subsec: feedback design} using the instruction: \textit{``Your refined feedback should be in a more human-like tone so that students are more willing to follow it.''} Second, the student's personalized cognitive state trajectory and the humanized feedback are jointly input, and the LLM contextualizes the message by incorporating recent cognitive state trends. Third, the contextualized feedback is processed by a Feedback Tone Translator module, which prompts the LLM to diversify tone while distinguishing it from previously delivered messages. Example tones include concise, friendly, high-energy, mandatory, gentle, playful/humorous, or formal/professional.  

A key constraint in refinement is that the LLM tutor must preserve the intended action (e.g., ``relax'' in relief feedback) and not alter its core directive. Importantly, the refinement process does not rely on course-related information, ensuring generalizability across tasks. Instead, it draws exclusively on sensing data from students' past cognitive state histories to contextualize feedback.

\section{Evaluation}
\label{sec: TutorLoop Evaluation}

We evaluate TutorLoop in a user study examining its impact on learning behaviors and outcomes.

\begin{figure*}
\centering
\includegraphics[width=1\linewidth]{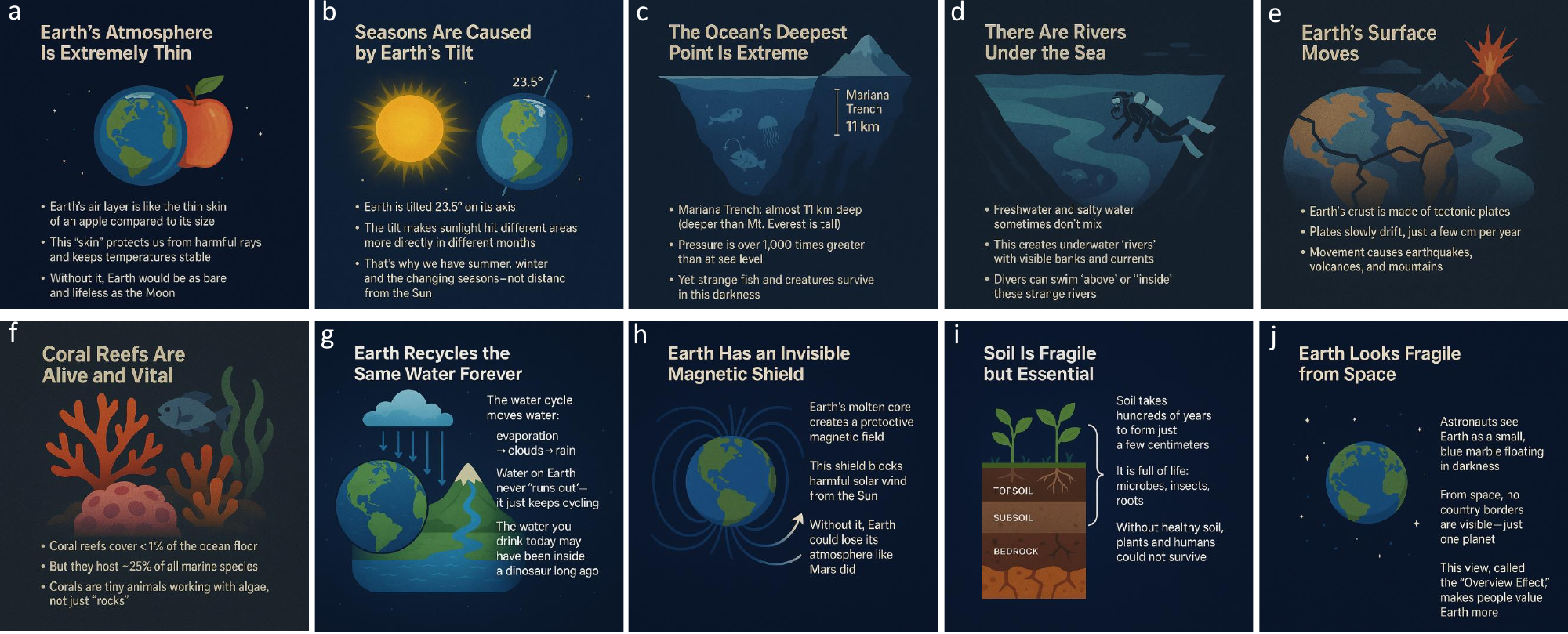}
\caption{
Ten slides in our learning task in the user study.
}
\Description{
A grid of ten slides shows the learning task materials. Each slide presents one concept about the Earth along with three supporting details. All slides (a–j) include visual illustrations and text descriptions such as geological processes, ecosystems, or atmospheric concepts. Each slide contains both text and images designed for comprehension.
}
\label{fig: slide material}
\end{figure*}

\subsection{Task}
\label{subsec:task}

\begin{figure}
\centering
\includegraphics[width=0.6\linewidth]{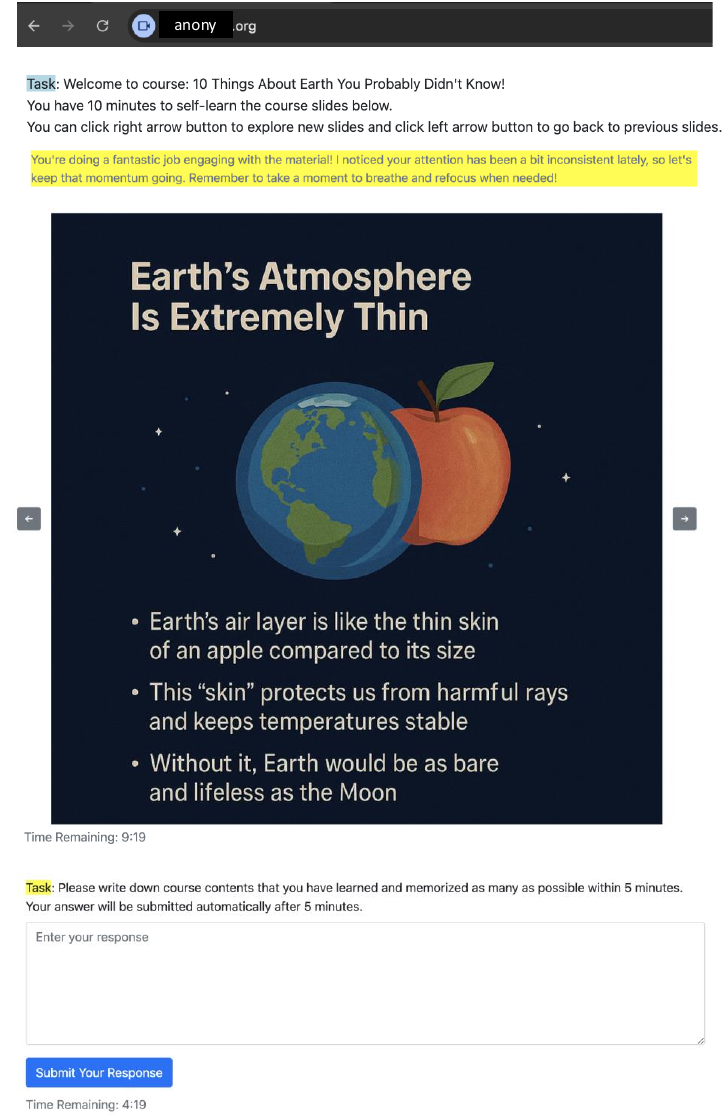}
\caption{
Task settings (learning session: top, testing session: bottom) in the user study.
}
\Description{
Screenshots of the user study platform. The top panel shows the learning session interface with a slide displayed at the center, navigation buttons left and right, a timer at the bottom, and feedback text appearing above the slide. The bottom panel shows the testing session interface with a blank text box for recalling concepts and details, also with a countdown timer displayed.
}
\label{fig: user study task setting}
\end{figure}

Our student simulator is based on the publicly available EduAgent dataset \cite{xu2024eduagent}, where the task involves students watching online course videos independently. To demonstrate the transferability of our approach beyond this dataset, we design a new self-learning task in which students study a set of ten slides unrelated to the original EduAgent materials. Students actively navigate the slides by clicking left and right buttons (Fig.~\ref{fig: user study task setting}), rather than passively viewing automatically played videos.

The slides, inspired by MinuteEarth \cite{minuteearth}, present ten unique concepts about the Earth (Fig.~\ref{fig: slide material}). Each slide introduces one core concept accompanied by three supporting details. This design allows us to quantitatively assess learning outcomes by measuring both the number of concepts recalled out of ten and the number of details remembered per concept.  

To develop the slides, we use GPT-4 to generate both textual and visual content. GPT-4 is prompted to produce ten slides, each with a unique concept and three supporting details, tailored to a teenage and young adult audience (ages 15–24) to ensure accessibility and comprehension. The final slide set is shown in Fig.~\ref{fig: slide material}.

The task comprises two sessions. In the \textbf{learning session}, participants study the slides within a fixed duration. In the subsequent \textbf{testing session}, participants are asked to write down, within a time limit, as many concepts and details as they can recall (Fig.~\ref{fig: user study task setting}). Learning outcomes are then evaluated by comparing participants' responses against the original slide content. Throughout sessions, the webcam remains active to capture cognitive states for analyzing learning behaviors.

\subsection{Study Design}

We adopted a between-subjects design with four groups: \textit{Control}, \textit{TutorUp}, \textit{DRL Only}, and \textit{Full}. All participants completed the task described in Section~\ref{subsec:task}. Groups differed only in the feedback received during the learning session; no feedback was provided during the testing session. Participants were randomly assigned to one of the four groups. In the \textit{Control} group, no feedback was delivered. In the \textit{Full} group, participants received feedback generated by TutorLoop, which integrates both the DRL model and LLM tutor. The DRL model was pre-trained in the virtual environment (the student simulator) with public EduAgent dataset, which provides cognitive states analogous to our setting, and was then directly applied to real participants without further updates. The other two groups serve as baselines, following prior work \cite{pan2025tutorup,bassen2020reinforcement}.

Specifically, the \textit{TutorUp} group adapts the approach from Pan et al.~\cite{pan2025tutorup}, which originally trained human tutors by interacting with simulated students to iteratively refine feedback to increase student engagement. In our adaptation, the LLM tutor directly refined its feedback with simulated students. For comparability with TutorLoop, we instantiate the simulated students using EduAgent dataset as well. Specifically, for each real participant, the LLM tutor retrieves similar virtual students from EduAgent via retrieval-augmented generation (RAG) and iterates its feedback generation before delivering it. Feedback generation follows the strategies in the original TutorUp framework, including either \textit{immediate feedback} (contextual and personalized) or \textit{asynchronous feedback} (longer and reflective), with the choice made adaptively by the LLM tutor based on real-time cognitive states. The tutor may also opt to deliver no feedback, mirroring our ``no feedback'' type. Additional implementation details follow the original TutorUp paper \cite{pan2025tutorup}.  

The \textit{DRL Only} group isolates the effect of the DRL model by removing the LLM tutor of TutorLoop. In this setting, only the unrefined versions of the four feedback types (Section~\ref{subsec: feedback design}) are delivered. This baseline serves two purposes. First, it functions as an ablation study to assess the added value of the LLM tutor by comparison with the \textit{Full} group. Second, it aligns with the work of Bassen et al.~\cite{bassen2020reinforcement}, who employed DRL to adaptively schedule educational activities. While our task differs, we adopt the same PPO architecture and adapt it to our feedback types, thereby providing a comparison to this prior line of research.

\subsection{Apparatus, Participants, and Procedures}

\textbf{Apparatus}: We developed a custom web-based learning platform for this study. During the learning session, the interface displayed one course slide at a time (Fig.~\ref{fig: user study task setting}). Participants navigated the ten slides using left and right buttons; each slide contained one key concept about the Earth and three supporting details (Section~\ref{subsec:task}). A timer at the bottom of the screen indicated the remaining session time. Feedback messages, when provided, appeared above the slide for 10 seconds (Fig.~\ref{fig: user study task setting}) before disappearing to minimize distraction. In the testing session, participants recalled and typed everything they had learned into a text box, with the remaining time again shown at the bottom of the screen.

\textbf{Participants}: Sample size was determined via a priori power analysis to achieve 80\% power to detect a medium effect size ($0.25$) at $\alpha=0.05$ in a four-group between-subjects design, yielding a target of $N=45$ per group. Participants were randomly assigned to groups until the minimum threshold was reached. Totally, we recruited 192 participants from Prolific.com, excluding five due to technical issues that caused the website to freeze, resulting in a final sample of 187 (Control: $N=48$, TutorUp: $N=46$, DRL Only: $N=46$, Full: $N=47$). The study was approved by our institutional IRB, and all participants provided informed consent and received compensation.

\begin{figure*}
\centering
\includegraphics[width=1\linewidth]{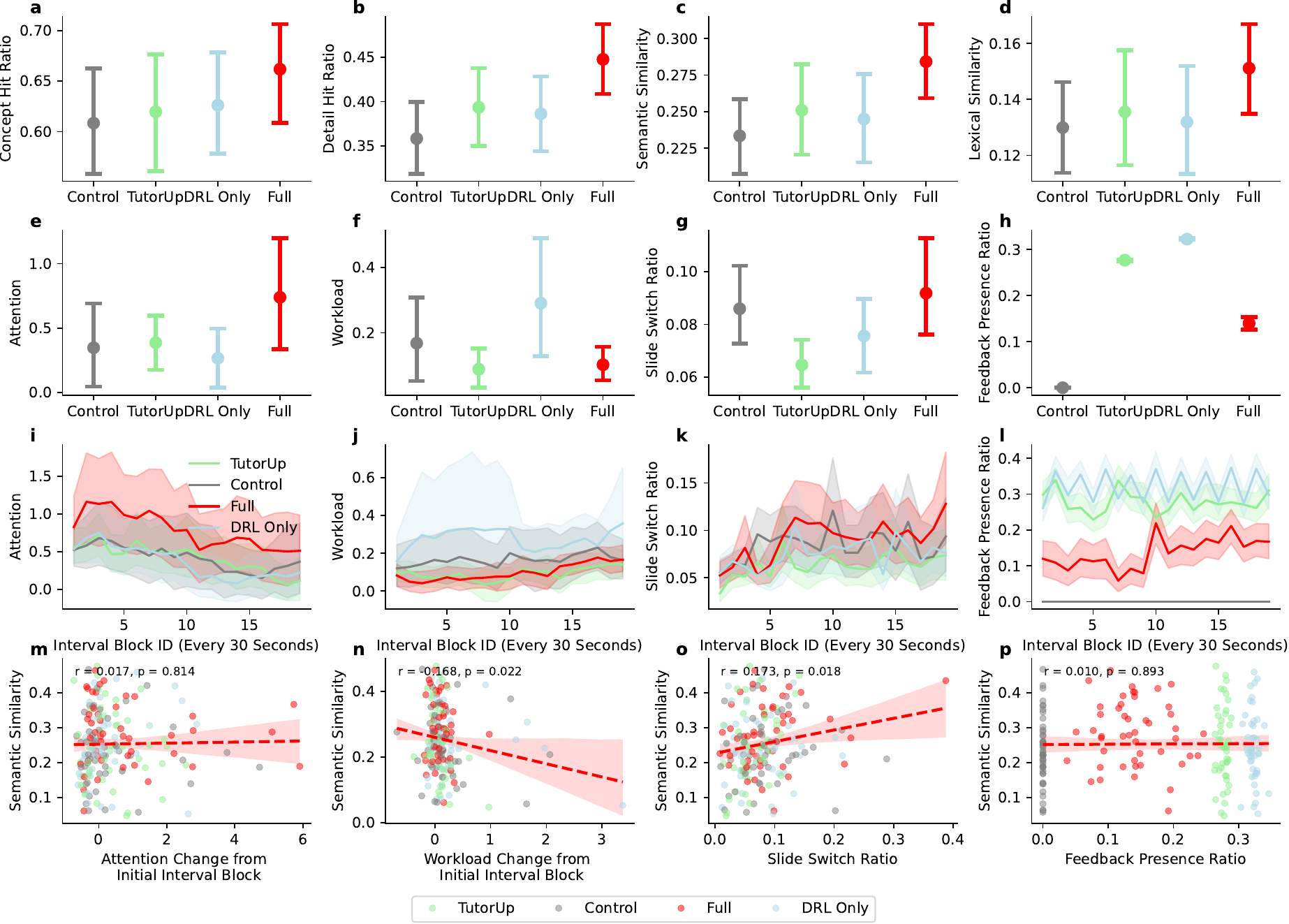}
\caption{
Results in our user study. 
}
\Description{
A multi-panel figure showing user study results. Bar plots compare Control, TutorUp, DRL Only, and Full groups across metrics. (a) Concept hit ratio: Control ~0.61, TutorUp ~0.62, DRL Only ~0.63, Full ~0.66. (b) Detail hit ratio: Control ~0.36, TutorUp ~0.39, DRL Only ~0.39, Full ~0.45. (c) Semantic similarity: Control ~0.23, TutorUp ~0.25, DRL Only ~0.25, Full ~0.28. (d) Lexical similarity: Control ~0.13, TutorUp ~0.14, DRL Only ~0.13, Full ~0.15. (e) Attention: Control ~0.35, TutorUp ~0.39, DRL Only ~0.27, Full ~0.74. (f) Workload: Control ~0.17, TutorUp ~0.09, DRL Only ~0.29, Full ~0.10. (g) Slide switch ratio: Control ~0.09, TutorUp ~0.06, DRL Only ~0.07, Full ~0.09. (h) Feedback presence ratio: Control 0, TutorUp ~0.28, DRL Only ~0.32, Full ~0.14. Line plots (i–l) show time trends of attention, workload, slide switching, and feedback presence over 15 interval blocks. Scatter plots (m–p) show correlations: attention vs. semantic similarity (r = 0.017), workload vs. similarity (r = −0.168), slide switching vs. similarity (r = 0.173), and feedback presence vs. similarity (r = 0.010).
}
\label{fig: user study results}
\end{figure*}

\textbf{Procedures}: The study began with an introduction and consent form, followed by task instructions. Prior to the learning session, participants completed the standard WebGazer calibration procedure\footnote{\url{https://webgazer.cs.brown.edu}} to account for individual differences and head movement. Specifically, they fixated and clicked nine calibration points across the screen five times. Participants proceeded only if eye-tracking accuracy exceeded 75\%. The learning session lasted 10 minutes, during which participants self-studied the slides. The subsequent testing session lasted 5 minutes, during which they recalled and wrote down as many concepts and details as possible. The total study duration was approximately 30 minutes. The 15-minute cap on the combined learning and testing sessions ensured high eye-tracking accuracy, as prior work \cite{xu2025peeredu} has shown WebGazer performance remains reliable for top 15 minutes post-calibration before drifting. The 10-minute learning duration is also consistent with prior studies employing short video-based materials such as MinuteEarth \cite{minuteearth} and MinutePhysics \cite{minutephysics} to assess student learning outcomes \cite{madsen2021synchronized,xu2025peeredu}.

\subsection{Evaluation Metrics and Measurements}

Attention and workload are recorded every second to capture fine-grained changes in cognitive states. Feedback, however, is generated once every 30 seconds (an interval block) and displayed for 10 seconds to minimize interruptions to learning. Below, we describe the evaluation metrics.

\subsubsection{Attention and Workload}
Following DRL reward design principles that account for individual differences \cite{schmeck1988individual}, we compute relative changes in attention and workload compared to each participant's initial baseline in the first 30-second block (before feedback). For interval block $i$, relative attention is defined as $Ua^{r}_{i} = (Ua_{i} - \hat{Ua})/\hat{Ua}$ and relative workload as $Uw^{r}_{i} = (Uw_{i} - \hat{Uw})/\hat{Uw}$, where $\hat{Ua}$ and $\hat{Uw}$ denote the initial baseline of attention and workload, respectively.

\subsubsection{Active Engagement}
As described in Section~\ref{subsec:task}, participants could switch slides using on-screen buttons. We measure active engagement through the \textit{slide switch ratio}, calculated as $S_r = m/600$, where $m$ is the number of button clicks during the 10-minute (600-second) learning session.

\subsubsection{Feedback Presence Ratio}
Since feedback strategies differ across groups, the duration of displayed feedback varies. We define the \textit{feedback presence ratio} as $F_r = t/600$, where $t$ is the total number of seconds feedback appeared in a 10-minute learning session. This metric can measure feedback generation efficiency to regulate student learning.

\subsubsection{Concept Hit Ratio}
To assess learning outcomes, we compare participants' written responses with the 10 key concepts in the course slides. If $k$ concepts are mentioned, the \textit{concept hit ratio} is $C_r = k/10$.

\subsubsection{Detail Hit Ratio}
As described in Section~\ref{subsec:task}, each slide includes three details (30 in total). If $k$ details are recalled in a participant's written response, the \textit{detail hit ratio} is $D_r = k/30$.

\subsubsection{Semantic Similarity}
We capture fine-grained alignment by embedding both written responses and course content using the OpenAI \texttt{text-embedding-small} model \cite{OpenAI} and then computing the cosine similarity between them.

\subsubsection{Lexical Similarity}
At the lexical level, we tokenize both written responses and course content into words and compute the overlap ratio as $L_r = \text{overlapped words} / \text{total words in course content}$, reflecting how much course content is reproduced at the word level.

\subsection{Results and Analysis}

We conduct statistical analyses using a linear mixed-effects model, which accounts for both group-level differences and individual-level variability across participants. This approach is appropriate as it is robust to non-normal data distributions.

\subsubsection{\textbf{Learning Behavioral Regulation}}

We analyze how feedback modulates learning behaviors across four dimensions: attention, workload, active engagement, and feedback presence.

\textit{\textbf{Attention}}. 
Table~\ref{tab:lme_overall_attention} reports results using the \textit{Full} group as the reference ($M = 0.739$), with pairwise comparisons in Table~\ref{tab:lme_pairwise_attention}. The \textit{Control} group exhibited significantly lower attention than \textit{Full} ($p = .013$), whereas differences between \textit{Control} and either \textit{TutorUp} or \textit{DRL Only} were not significant. Moreover, the \textit{Full} group significantly outperformed both \textit{TutorUp} ($p = .044$) and \textit{DRL Only} ($p = .002$), highlighting the benefit of integrating DRL with LLM-based refinement. 
Figure~\ref{fig: user study results}(i) further shows that attention in the \textit{Full} group remains consistently higher across most time blocks, indicating robust and sustained improvements.

\textit{\textbf{Workload}}.  
Table~\ref{tab:lme_overall_workload} (with pairwise comparisons in Table~\ref{tab:lme_pairwise_workload}) shows that the \textit{Full} group significantly reduced workload relative to \textit{DRL Only} ($p < .001$). The \textit{TutorUp} group also achieved lower workload than \textit{DRL Only} ($p = .017$), and was comparable to \textit{Full}. Although neither \textit{Full} nor \textit{TutorUp} significantly differed from \textit{Control}, both exhibit consistently lower workload trends over time (Fig.~\ref{fig: user study results}(j)). 
These results suggest that LLM-based feedback refinement plays a key role in mitigating workload, while purely DRL-driven strategies may introduce unnecessary cognitive burden.

\textit{\textbf{Active Engagement}}.  
Table~\ref{tab:lme_overall_slide_switch} (pairwise results in Table~\ref{tab:lme_pairwise_slide_switch}) reports active engagement (slide switch ratio) with \textit{Full} as the reference ($M = 0.092$). Unlike attention, both \textit{TutorUp} and \textit{DRL Only} groups showed reduced engagement relative to \textit{Control}, with a significant decrease for \textit{TutorUp} ($p = .040$). In contrast, the \textit{Full} group maintained slightly higher engagement than \textit{Control} (n.s.) and significantly outperformed both \textit{TutorUp} and \textit{DRL Only} ($p \leq .002$).
This pattern suggests that improvements in attention do not necessarily translate into active engagement. Baseline approaches may increase focus at the expense of interaction, whereas the integrated \textit{Full} approach better balances both. Figure~\ref{fig: user study results}(k) corroborates this trend, showing consistently higher engagement in the \textit{Full} group.

\textit{\textbf{Feedback Presence Ratio}}.
Table~\ref{tab:lme_overall_feedback_presence} (pairwise results in Table~\ref{tab:lme_pairwise_feedback_presence}) shows that all treatment groups delivered more feedback than \textit{Control}, as expected. Notably, \textit{DRL Only} produced significantly more feedback than \textit{TutorUp} ($p < .001$), indicating inefficient over-generation. In contrast, the \textit{Full} group significantly reduced feedback frequency compared to both \textit{DRL Only} and \textit{TutorUp} (both $p < .001$), suggesting more selective and efficient intervention.
Temporal analysis (Fig.~\ref{fig: user study results}(l)) reveals that the \textit{Full} group delivers less feedback early and more later in the session, in contrast to the near-constant patterns in baseline conditions. This adaptive strategy aligns with observed behavioral dynamics: attention decreases and workload increases over time (Fig.~\ref{fig: user study results}(i,j)). 

Importantly, although both \textit{Full} and \textit{DRL Only} employ DRL (PPO), the \textit{Full} model incorporates LLM-refined feedback during training, whereas \textit{DRL Only} relies solely on a student simulator. This integration enables the policy to more effectively select ``no feedback'' when appropriate, reducing unnecessary interventions while maintaining learning effectiveness.

Overall, these results demonstrate that our integrated approach not only improves attention, but also better regulates workload, sustains engagement, and adaptively controls feedback delivery in response to students’ evolving cognitive states.

\subsubsection{\textbf{Learning Outcome Improvement}}

We analyze how feedback modulates learning outcomes across groups using four complementary metrics: concept hit ratio, detail hit ratio, semantic similarity, and lexical similarity (Tables~\ref{tab:lme_overall_concept_hit_ratio}–\ref{tab:lme_overall_lexical}). Across all metrics, the \textit{Full} group consistently achieved the highest performance (concept: $M = 0.662$; detail: $M = 0.448$; semantic: $M = 0.284$; lexical: $M = 0.151$).

Compared to the \textit{Control} group, all three treatment groups showed improved learning outcomes. However, only the \textit{Full} condition yielded statistically significant gains across all four metrics (concept: $p = .031$; detail: $p < .001$; semantic: $p < .001$; lexical: $p = .029$), highlighting the advantage of the integrated approach.

When comparing among treatment conditions, a more nuanced pattern emerges. For higher-level understanding (concept hit ratio), the \textit{Full} group showed improvements over both \textit{DRL Only} and \textit{TutorUp}, though these differences were not statistically significant. In contrast, for more fine-grained measures (detail hit ratio and semantic similarity), the \textit{Full} group significantly outperformed both alternatives ($p < .01$), suggesting stronger support for retaining detailed information and capturing meaning. For lexical similarity, the \textit{Full} group significantly outperformed \textit{DRL Only} ($p = .049$), but not \textit{TutorUp}, likely due to variation in phrasing despite similar underlying semantics.

Taken together, these results suggest that while all feedback mechanisms improve learning outcomes, the integrated \textit{Full} approach is particularly effective for promoting deeper and more detailed learning, beyond high-level concept acquisition.

\subsubsection{\textbf{Correlation between Learning Behaviors and Learning Outcomes}}

Using semantic similarity to represent learning outcomes provides additional insights for the correlation with learning behaviors, as shown in Fig.~\ref{fig: user study results}(m,n,o,p).  


\textbf{Beyond attention regulation.}
Although attention was only weakly correlated with learning outcomes ($r = 0.017$, $p = 0.814$), workload showed a strong negative correlation ($r = -0.168$, $p = 0.022$). This suggests that simply delivering feedback to increase students' visual attention may not substantially improve learning outcomes, and that naïvely using LLM feedback to raise attention could inadvertently introduce cognitive overload. These findings underscore the importance of our accumulated reward optimization, which balances attention and workload in the design of feedback.  

\textbf{Active engagement.}
It is also notable that feedback content in all groups focused only on either increasing attention or encouraging rest to reduce workload, and did not include explicit prompts to increase engagement through slide switching. Nevertheless, the significant improvement in active engagement (slide switching) in our \textit{Full} group, combined with the positive correlation between active engagement and learning outcomes ($r = 0.173$, $p = 0.018$), suggests that our system can enhance learning by fostering active engagement implicitly, since students were never directly instructed to do so.  

\textbf{Feedback frequency.}
Finally, we observed a near-zero correlation between feedback presence ratio and learning outcomes ($r = 0.010$, $p = 0.893$). This indicates that more frequent feedback does not necessarily lead to better outcomes. Taken together with earlier results showing that less frequent feedback in our model yielded stronger learning outcomes, these findings demonstrate the effectiveness of adaptive feedback. Rather than providing constant feedback, which can increase workload during learning, adaptive and targeted feedback appears to be more beneficial than frequent and intrusive interventions.

\section{Discussion}

\subsection{Implications for Designing Feedback Systems to Augment Student Learning}


A key implication for HCI researchers lies in how simulators and hybrid AI models can be integrated to design feedback systems at scale. The student simulator allowed us to pre-train reinforcement learning policies on simulated cognitive trajectories, making it possible to optimize long-term strategies without costly human-in-the-loop data collection. At the same time, our results show that a pure DRL model tends to over-deliver feedback and raise workload, while the combined DRL+LLM design achieved greater efficiency and effectiveness. This points toward a productive division of labor: simulators and DRL agents should be tasked with long-horizon optimization, such as deciding whether and when feedback should be delivered, while LLMs should focus on near-term communicative goals, such as phrasing, tone, and personalization. For HCI research, this hybrid approach offers both scalability and human-centeredness: simulators enable generalizable optimization across tasks, while LLMs ensure that interventions remain contextually sensitive and socially acceptable when deployed in real-world learning environments.


Our findings reveal that attention alone is not a reliable predictor of learning success, whereas workload shows a strong negative correlation with outcomes. This suggests that designs narrowly focused on maximizing attention may backfire by increasing workload and diminishing performance. For HCI researchers, this highlights the importance of treating attention and workload as two sides of a cognitive tradeoff rather than independent targets. Feedback systems should therefore be designed to regulate the \emph{balance} between attention and workload, guided by the Yerkes–Dodson law. Practically, this requires building interfaces that detect both over-engagement (when learners are pushing beyond their optimal workload) and under-engagement (when attention drifts) and adapt feedback accordingly. Designing for equilibrium, rather than maximization, ensures that feedback is not just stimulating but also sustainable across longer learning sessions.

The superior performance of our \textbf{TutorLoop} system, which delivered fewer yet more strategically timed interventions, demonstrates that the \emph{timing} of feedback matters more than its frequency. Over-delivery of prompts, as seen in the \textbf{DRL Only} group, elevated workload and impaired learning outcomes. For HCI researchers, this underscores the need to move away from rigid or high-frequency feedback schedules. Instead, adaptive timing strategies that align with learners' trajectories should be prioritized, for instance, intervening during moments of declining attention or rising workload rather than at fixed intervals. This principle also resonates with broader HCI concerns around notification design \cite{reza2023exam}, where poorly timed interventions can be disruptive and counterproductive \cite{walker2003heading}. By integrating models that learn to detect and anticipate cognitive shifts, HCI systems can deliver fewer but more impactful interventions, reducing cognitive intrusiveness while maximizing benefit.

Finally, our framework demonstrates that grounding feedback in generalizable signals, such as attention and workload derived from sensors, enables direct transfer across learning contexts without retraining. This portability contrasts with many prior systems that rely on course-specific content or domain-specific knowledge, which limits scalability. For HCI researchers, the implication is that feedback systems should increasingly leverage domain-agnostic behavioral signals (e.g., gaze, facial micro-movements, physiological workload indicators) to design interventions that are flexible across subjects, platforms, and learner populations. Portability not only reduces deployment costs but also raises important design opportunities: feedback systems could be embedded across diverse online learning environments, from short-form microlearning apps to longer MOOCs, without extensive customization. Such portability expands the potential reach of HCI-designed interventions, while also posing challenges around privacy, ethical sensing, and cross-context calibration that future work should address.

\subsection{Limitations and Future Work}

One limitation lies in the duration of our learning task (10 minutes). While this design choice is consistent with prior studies that use short educational videos such as Minute Earth \cite{minuteearth} and Minute Physics \cite{minutephysics} to assess learning performance \cite{madsen2021synchronized,xu2025peeredu}, it restricts our ability to observe longer-term dynamics. Extending the task duration in future work may reveal new insights into how adaptive feedback influences learning over extended periods.  

A second limitation concerns our operationalization of arousal in the Yerkes–Dodson law \cite{yerkes1908relation}. We used visual attention as a proxy for arousal due to its strong empirical correlation \cite{janelle2002anxiety}, yet the two constructs are not identical. More direct measures such as EEG could provide richer information about arousal, though their limited scalability makes them impractical for remote online learning platforms. In contrast, web cameras offer a more feasible compromise between accuracy and accessibility.  

Third, our evaluation of learning outcomes relied primarily on working memory, as measured by recall in the self-learning task. While working memory is a critical component of cognition and learning \cite{cowan2014working}, it does not capture the full spectrum of learning performance \cite{sfard2001cognition}. Broader indicators, such as comprehension assessed through post-tests \cite{xu2025peeredu}, could provide complementary evidence. Future research should therefore examine a wider range of cognitive functions to build a more comprehensive picture of learning outcomes.

\section{Conclusion}

This work introduces \textbf{TutorLoop}, a sensor-in-the-loop framework that integrates reinforcement learning and large language models to deliver adaptive feedback grounded in cognitive states. Across a large-scale study with 187 participants, TutorLoop significantly improved attention, reduced workload, fostered active engagement, and enhanced learning outcomes compared to both prior approaches and ablated baselines. Our findings highlight that effective feedback design requires balancing attention and workload, delivering interventions adaptively rather than frequently. By grounding feedback in sensor-derived states rather than course-specific content, TutorLoop demonstrates promising transferability across tasks and contexts, offering a path toward scalable, domain-agnostic learning support. More broadly, this work suggests that HCI researchers should reconceptualize feedback not as static prompts but as adaptive, theory-guided interactions that integrate algorithmic optimization with human-centered communication to augment student learning.

\bibliographystyle{ACM-Reference-Format}
\bibliography{0_main}

\newpage

\appendix

\section{Appendix: details in experimental results}

\begin{table}[H]
\centering
\caption{Statistical Results (Attention)}
\Description{
Statistical results for attention. Intercept = 0.739 (SE 0.015, z = 50.036, p = 0.000, CI [0.710, 0.768]). TutorUp = −0.353 (SE 0.175, z = −2.010, p = 0.044, CI [−0.697, −0.009]). Control = −0.392 (SE 0.159, z = −2.471, p = 0.013, CI [−0.703, −0.081]). DRL Only = −0.472 (SE 0.156, z = −3.037, p = 0.002, CI [−0.777, −0.168]).
}
\label{tab:lme_overall_attention}
\begin{tabular}{lcccccc}
\hline
\textbf{Group} & \textbf{Coef.} & \textbf{Std.Err.} & \textbf{z} & \textbf{P>|z|} & \textbf{[0.025} & \textbf{0.975]} \\
\hline
Intercept & 0.739 & 0.015 & 50.036 & 0.000 & 0.710 & 0.768 \\
TutorUp & -0.353 & 0.175 & -2.010 & 0.044 & -0.697 & -0.009 \\
Control & -0.392 & 0.159 & -2.471 & 0.013 & -0.703 & -0.081 \\
DRL Only & -0.472 & 0.156 & -3.037 & 0.002 & -0.777 & -0.168 \\
\hline
\end{tabular}
\end{table}

\FloatBarrier

\begin{table}[H]
\centering
\caption{Pairwise Comparisons (Attention)}
\Description{
Pairwise comparisons for attention. Control vs DRL Only = 0.081 (SE 0.229, z = 0.352, p = 0.725, CI [−0.369, 0.530]). Control vs Full = −0.392 (SE 0.159, z = −2.471, p = 0.013, CI [−0.703, −0.081]). Control vs TutorUp = −0.039 (SE 0.229, z = −0.171, p = 0.864, CI [−0.488, 0.410]). DRL Only vs Full = −0.473 (SE 0.156, z = −3.037, p = 0.002, CI [−0.777, −0.168]). DRL Only vs TutorUp = −0.120 (SE 0.231, z = −0.519, p = 0.604, CI [−0.572, 0.333]). Full vs TutorUp = 0.353 (SE 0.175, z = 2.010, p = 0.044, CI [0.009, 0.697]).
}
\label{tab:lme_pairwise_attention}
\setlength{\tabcolsep}{2pt}
\begin{tabular}{lcccccc}
\hline
\textbf{Comparison} & \textbf{Coef.} & \textbf{Std.Err.} & \textbf{z} & \textbf{P>|z|} & \textbf{[0.025} & \textbf{0.975]} \\
\hline
Control vs DRL Only & 0.081 & 0.229 & 0.352 & 0.725 & -0.369 & 0.530 \\
Control vs Full & -0.392 & 0.159 & -2.471 & 0.013 & -0.703 & -0.081 \\
Control vs TutorUp & -0.039 & 0.229 & -0.171 & 0.864 & -0.488 & 0.410 \\
DRL Only vs Full & -0.473 & 0.156 & -3.037 & 0.002 & -0.777 & -0.168 \\
DRL Only vs TutorUp & -0.120 & 0.231 & -0.519 & 0.604 & -0.572 & 0.333 \\
Full vs TutorUp & 0.353 & 0.175 & 2.010 & 0.044 & 0.009 & 0.697 \\
\hline
\end{tabular}
\end{table}

\FloatBarrier

\begin{table}[H]
\centering
\caption{Statistical Results (Workload)}
\Description{
Statistical results for workload. Intercept = 0.102 (SE 0.028, z = 3.627, p = 0.000, CI [0.047, 0.158]). TutorUp = −0.014 (SE 0.059, z = −0.232, p = 0.817, CI [−0.130, 0.102]). Control = 0.066 (SE 0.039, z = 1.674, p = 0.094, CI [−0.011, 0.143]). DRL Only = 0.188 (SE 0.049, z = 3.801, p = 0.000, CI [0.091, 0.285]).
}
\label{tab:lme_overall_workload}
\begin{tabular}{lcccccc}
\hline
\textbf{Group} & \textbf{Coef.} & \textbf{Std.Err.} & \textbf{z} & \textbf{P>|z|} & \textbf{[0.025} & \textbf{0.975]} \\
\hline
Intercept & 0.102 & 0.028 & 3.627 & 0.000 & 0.047 & 0.158 \\
TutorUp & -0.014 & 0.059 & -0.232 & 0.817 & -0.130 & 0.102 \\
Control & 0.066 & 0.039 & 1.674 & 0.094 & -0.011 & 0.143 \\
DRL Only & 0.188 & 0.049 & 3.801 & 0.000 & 0.091 & 0.285 \\
\hline
\end{tabular}
\end{table}

\FloatBarrier

\begin{table}[H]
\centering
\caption{Pairwise Comparisons (Workload)}
\Description{
Pairwise comparisons for workload. Control vs DRL Only = −0.122 (SE 0.084, z = −1.454, p = 0.146, CI [−0.287, 0.043]). Control vs Full = 0.066 (SE 0.039, z = 1.674, p = 0.094, CI [−0.011, 0.143]). Control vs TutorUp = 0.080 (SE 0.083, z = 0.957, p = 0.339, CI [−0.083, 0.242]). DRL Only vs Full = 0.188 (SE 0.049, z = 3.801, p = 0.000, CI [0.091, 0.285]). DRL Only vs TutorUp = 0.202 (SE 0.085, z = 2.381, p = 0.017, CI [0.036, 0.368]). Full vs TutorUp = 0.014 (SE 0.059, z = 0.232, p = 0.817, CI [−0.102, 0.130]).
}
\label{tab:lme_pairwise_workload}
\setlength{\tabcolsep}{2pt}
\begin{tabular}{lcccccc}
\hline
\textbf{Comparison} & \textbf{Coef.} & \textbf{Std.Err.} & \textbf{z} & \textbf{P>|z|} & \textbf{[0.025} & \textbf{0.975]} \\
\hline
Control vs DRL Only & -0.122 & 0.084 & -1.454 & 0.146 & -0.287 & 0.043 \\
Control vs Full & 0.066 & 0.039 & 1.674 & 0.094 & -0.011 & 0.143 \\
Control vs TutorUp & 0.080 & 0.083 & 0.957 & 0.339 & -0.083 & 0.242 \\
DRL Only vs Full & 0.188 & 0.049 & 3.801 & 0.000 & 0.091 & 0.285 \\
DRL Only vs TutorUp & 0.202 & 0.085 & 2.381 & 0.017 & 0.036 & 0.368 \\
Full vs TutorUp & 0.014 & 0.059 & 0.232 & 0.817 & -0.102 & 0.130 \\
\hline
\end{tabular}
\end{table}
\FloatBarrier

\begin{table}[H]
\centering
\caption{Statistical Results (Slide Switch Ratio)}
\Description{
Statistical results for slide switch ratio. Intercept = 0.092 (SE 0.002, z = 43.045, p = 0.000, CI [0.088, 0.096]). TutorUp = −0.027 (SE 0.008, z = −3.434, p = 0.001, CI [−0.043, −0.012]). Control = −0.006 (SE 0.008, z = −0.714, p = 0.475, CI [−0.022, 0.010]). DRL Only = −0.016 (SE 0.005, z = −3.072, p = 0.002, CI [−0.027, −0.006]).
}
\label{tab:lme_overall_slide_switch}
\begin{tabular}{lcccccc}
\hline
\textbf{Group} & \textbf{Coef.} & \textbf{Std.Err.} & \textbf{z} & \textbf{P>|z|} & \textbf{[0.025} & \textbf{0.975]} \\
\hline
Intercept & 0.092 & 0.002 & 43.045 & 0.000 & 0.088 & 0.096 \\
TutorUp & -0.027 & 0.008 & -3.434 & 0.001 & -0.043 & -0.012 \\
Control & -0.006 & 0.008 & -0.714 & 0.475 & -0.022 & 0.010 \\
DRL Only & -0.016 & 0.005 & -3.072 & 0.002 & -0.027 & -0.006 \\
\hline
\end{tabular}
\end{table}

\FloatBarrier

\begin{table}[H]
\centering
\caption{Pairwise Comparisons (Slide Switch Ratio)}
\Description{
Pairwise comparisons for slide switch ratio. Control vs DRL Only = 0.010 (SE 0.010, z = 1.031, p = 0.303, CI [−0.009, 0.030]). Control vs Full = −0.006 (SE 0.008, z = −0.714, p = 0.475, CI [−0.022, 0.010]). Control vs TutorUp = 0.021 (SE 0.010, z = 2.049, p = 0.040, CI [0.001, 0.042]). DRL Only vs Full = −0.016 (SE 0.005, z = −3.072, p = 0.002, CI [−0.027, −0.006]). DRL Only vs TutorUp = 0.011 (SE 0.010, z = 1.065, p = 0.287, CI [−0.009, 0.031]). Full vs TutorUp = 0.027 (SE 0.008, z = 3.434, p = 0.001, CI [0.012, 0.043]).
}
\label{tab:lme_pairwise_slide_switch}
\setlength{\tabcolsep}{2pt}
\begin{tabular}{lcccccc}
\hline
\textbf{Comparison} & \textbf{Coef.} & \textbf{Std.Err.} & \textbf{z} & \textbf{P>|z|} & \textbf{[0.025} & \textbf{0.975]} \\
\hline
Control vs DRL Only & 0.010 & 0.010 & 1.031 & 0.303 & -0.009 & 0.030 \\
Control vs Full & -0.006 & 0.008 & -0.714 & 0.475 & -0.022 & 0.010 \\
Control vs TutorUp & 0.021 & 0.010 & 2.049 & 0.040 & 0.001 & 0.042 \\
DRL Only vs Full & -0.016 & 0.005 & -3.072 & 0.002 & -0.027 & -0.006 \\
DRL Only vs TutorUp & 0.011 & 0.010 & 1.065 & 0.287 & -0.009 & 0.031 \\
Full vs TutorUp & 0.027 & 0.008 & 3.434 & 0.001 & 0.012 & 0.043 \\
\hline
\end{tabular}
\end{table}
\FloatBarrier

\begin{table}[H]
\centering
\caption{Statistical Results (Feedback Presence Ratio)}
\Description{
Statistical results for feedback presence ratio. Intercept = 0.140 (SE 0.001, z = 152.196, p = 0.000, CI [0.138, 0.142]). TutorUp = 0.137 (SE 0.004, z = 32.302, p = 0.000, CI [0.129, 0.145]). Control = −0.140 (SE 0.004, z = −37.283, p = 0.000, CI [−0.147, −0.132]). DRL Only = 0.183 (SE 0.003, z = 62.697, p = 0.000, CI [0.177, 0.188]).
}
\label{tab:lme_overall_feedback_presence}
\begin{tabular}{lcccccc}
\hline
\textbf{Group} & \textbf{Coef.} & \textbf{Std.Err.} & \textbf{z} & \textbf{P>|z|} & \textbf{[0.025} & \textbf{0.975]} \\
\hline
Intercept & 0.140 & 0.001 & 152.196 & 0.000 & 0.138 & 0.142 \\
TutorUp & 0.137 & 0.004 & 32.302 & 0.000 & 0.129 & 0.145 \\
Control & -0.140 & 0.004 & -37.283 & 0.000 & -0.147 & -0.132 \\
DRL Only & 0.183 & 0.003 & 62.697 & 0.000 & 0.177 & 0.188 \\
\hline
\end{tabular}
\end{table}
\FloatBarrier

\begin{table}[H]
\centering
\caption{Pairwise Comparisons (Feedback Presence Ratio)}
\Description{
Pairwise comparisons for feedback presence ratio. Control vs DRL Only = −0.322 (SE 0.005, z = −62.763, p = 0.000, CI [−0.332, −0.312]). Control vs Full = −0.140 (SE 0.004, z = −37.283, p = 0.000, CI [−0.147, −0.132]). Control vs TutorUp = −0.277 (SE 0.005, z = −53.511, p = 0.000, CI [−0.287, −0.267]). DRL Only vs Full = 0.183 (SE 0.003, z = 62.697, p = 0.000, CI [0.177, 0.188]). DRL Only vs TutorUp = 0.046 (SE 0.005, z = 8.978, p = 0.000, CI [0.036, 0.056]). Full vs TutorUp = −0.137 (SE 0.004, z = −32.302, p = 0.000, CI [−0.145, −0.129]).
}
\label{tab:lme_pairwise_feedback_presence}
\setlength{\tabcolsep}{2pt}
\begin{tabular}{lcccccc}
\hline
\textbf{Comparison} & \textbf{Coef.} & \textbf{Std.Err.} & \textbf{z} & \textbf{P>|z|} & \textbf{[0.025} & \textbf{0.975]} \\
\hline
Control vs DRL Only & -0.322 & 0.005 & -62.763 & 0.000 & -0.332 & -0.312 \\
Control vs Full & -0.140 & 0.004 & -37.283 & 0.000 & -0.147 & -0.132 \\
Control vs TutorUp & -0.277 & 0.005 & -53.511 & 0.000 & -0.287 & -0.267 \\
DRL Only vs Full & 0.183 & 0.003 & 62.697 & 0.000 & 0.177 & 0.188 \\
DRL Only vs TutorUp & 0.046 & 0.005 & 8.978 & 0.000 & 0.036 & 0.056 \\
Full vs TutorUp & -0.137 & 0.004 & -32.302 & 0.000 & -0.145 & -0.129 \\
\hline
\end{tabular}
\end{table}
\FloatBarrier

\begin{table}[H]
\centering
\caption{Statistical Results (Concept Hit Ratio)}
\Description{
Statistical results for concept hit ratio. Intercept = 0.662 (SE 0.006, z = 111.460, p = 0.000, CI [0.650, 0.673]). TutorUp = −0.042 (SE 0.028, z = −1.494, p = 0.135, CI [−0.097, 0.013]). Control = −0.053 (SE 0.025, z = −2.153, p = 0.031, CI [−0.102, −0.005]). DRL Only = −0.036 (SE 0.021, z = −1.697, p = 0.090, CI [−0.077, 0.006]).
}
\label{tab:lme_overall_concept_hit_ratio}
\begin{tabular}{lcccccc}
\hline
\textbf{Group} & \textbf{Coef.} & \textbf{Std.Err.} & \textbf{z} & \textbf{P>|z|} & \textbf{[0.025} & \textbf{0.975]} \\
\hline
Intercept & 0.662 & 0.006 & 111.460 & 0.000 & 0.650 & 0.673 \\
TutorUp & -0.042 & 0.028 & -1.494 & 0.135 & -0.097 & 0.013 \\
Control & -0.053 & 0.025 & -2.153 & 0.031 & -0.102 & -0.005 \\
DRL Only & -0.036 & 0.021 & -1.697 & 0.090 & -0.077 & 0.006 \\
\hline
\end{tabular}
\end{table}
\FloatBarrier

\begin{table}[H]
\centering
\caption{Pairwise Comparisons (Concept Hit Ratio)}
\Description{
Pairwise comparisons for concept hit ratio. Control vs DRL Only = −0.018 (SE 0.038, z = −0.473, p = 0.636, CI [−0.091, 0.056]). Control vs Full = −0.053 (SE 0.025, z = −2.153, p = 0.031, CI [−0.102, −0.005]). Control vs TutorUp = −0.011 (SE 0.038, z = −0.299, p = 0.765, CI [−0.085, 0.062]). DRL Only vs Full = −0.036 (SE 0.021, z = −1.697, p = 0.090, CI [−0.077, 0.006]). DRL Only vs TutorUp = 0.007 (SE 0.038, z = 0.174, p = 0.862, CI [−0.067, 0.080]). Full vs TutorUp = 0.042 (SE 0.028, z = 1.494, p = 0.135, CI [−0.013, 0.097]).
}
\label{tab:lme_pairwise_concept_hit_ratio}
\setlength{\tabcolsep}{2pt}
\begin{tabular}{lcccccc}
\hline
\textbf{Comparison} & \textbf{Coef.} & \textbf{Std.Err.} & \textbf{z} & \textbf{P>|z|} & \textbf{[0.025} & \textbf{0.975]} \\
\hline
Control vs DRL Only & -0.018 & 0.038 & -0.473 & 0.636 & -0.091 & 0.056 \\
Control vs Full & -0.053 & 0.025 & -2.153 & 0.031 & -0.102 & -0.005 \\
Control vs TutorUp & -0.011 & 0.038 & -0.299 & 0.765 & -0.085 & 0.062 \\
DRL Only vs Full & -0.036 & 0.021 & -1.697 & 0.090 & -0.077 & 0.006 \\
DRL Only vs TutorUp & 0.007 & 0.038 & 0.174 & 0.862 & -0.067 & 0.080 \\
Full vs TutorUp & 0.042 & 0.028 & 1.494 & 0.135 & -0.013 & 0.097 \\
\hline
\end{tabular}
\end{table}
\FloatBarrier

\begin{table}[H]
\centering
\caption{Statistical Results (Detail Hit Ratio)}
\Description{
Statistical results for detail hit ratio. Intercept = 0.448 (SE 0.001, z = 356.892, p = 0.000, CI [0.445, 0.450]). TutorUp = −0.054 (SE 0.021, z = −2.634, p = 0.008, CI [−0.094, −0.014]). Control = −0.089 (SE 0.020, z = −4.416, p = 0.000, CI [−0.129, −0.050]). DRL Only = −0.061 (SE 0.020, z = −3.069, p = 0.002, CI [−0.100, −0.022]).
}
\label{tab:lme_overall_detail_hit_ratio}
\begin{tabular}{lcccccc}
\hline
\textbf{Group} & \textbf{Coef.} & \textbf{Std.Err.} & \textbf{z} & \textbf{P>|z|} & \textbf{[0.025} & \textbf{0.975]} \\
\hline
Intercept & 0.448 & 0.001 & 356.892 & 0.000 & 0.445 & 0.450 \\
TutorUp & -0.054 & 0.021 & -2.634 & 0.008 & -0.094 & -0.014 \\
Control & -0.089 & 0.020 & -4.416 & 0.000 & -0.129 & -0.050 \\
DRL Only & -0.061 & 0.020 & -3.069 & 0.002 & -0.100 & -0.022 \\
\hline
\end{tabular}
\end{table}

\FloatBarrier

\begin{table}[H]
\centering
\caption{Pairwise Comparisons (Detail Hit Ratio)}
\Description{
Pairwise comparisons for detail hit ratio. Control vs DRL Only = −0.028 (SE 0.029, z = −0.948, p = 0.343, CI [−0.086, 0.030]). Control vs Full = −0.089 (SE 0.020, z = −4.416, p = 0.000, CI [−0.129, −0.050]). Control vs TutorUp = −0.035 (SE 0.029, z = −1.194, p = 0.232, CI [−0.093, 0.023]). DRL Only vs Full = −0.061 (SE 0.020, z = −3.069, p = 0.002, CI [−0.100, −0.022]). DRL Only vs TutorUp = −0.007 (SE 0.030, z = −0.244, p = 0.808, CI [−0.066, 0.051]). Full vs TutorUp = 0.054 (SE 0.021, z = 2.634, p = 0.008, CI [0.014, 0.094]).
}
\label{tab:lme_pairwise_detail_hit_ratio}
\setlength{\tabcolsep}{2pt}
\begin{tabular}{lcccccc}
\hline
\textbf{Comparison} & \textbf{Coef.} & \textbf{Std.Err.} & \textbf{z} & \textbf{P>|z|} & \textbf{[0.025} & \textbf{0.975]} \\
\hline
Control vs DRL Only & -0.028 & 0.029 & -0.948 & 0.343 & -0.086 & 0.030 \\
Control vs Full & -0.089 & 0.020 & -4.416 & 0.000 & -0.129 & -0.050 \\
Control vs TutorUp & -0.035 & 0.029 & -1.194 & 0.232 & -0.093 & 0.023 \\
DRL Only vs Full & -0.061 & 0.020 & -3.069 & 0.002 & -0.100 & -0.022 \\
DRL Only vs TutorUp & -0.007 & 0.030 & -0.244 & 0.808 & -0.066 & 0.051 \\
Full vs TutorUp & 0.054 & 0.021 & 2.634 & 0.008 & 0.014 & 0.094 \\
\hline
\end{tabular}
\end{table}
\FloatBarrier

\begin{table}[H]
\centering
\caption{Statistical Results (Semantic Similarity)}
\Description{
Statistical results for semantic similarity. Intercept = 0.284 (SE 0.003, z = 94.678, p = 0.000, CI [0.278, 0.290]). TutorUp = −0.033 (SE 0.012, z = −2.769, p = 0.006, CI [−0.057, −0.010]). Control = −0.051 (SE 0.013, z = −4.052, p = 0.000, CI [−0.075, −0.026]). DRL Only = −0.039 (SE 0.014, z = −2.768, p = 0.006, CI [−0.067, −0.011]).
}
\label{tab:lme_overall_semantic}
\begin{tabular}{lcccccc}
\hline
\textbf{Group} & \textbf{Coef.} & \textbf{Std.Err.} & \textbf{z} & \textbf{P>|z|} & \textbf{[0.025} & \textbf{0.975]} \\
\hline
Intercept & 0.284 & 0.003 & 94.678 & 0.000 & 0.278 & 0.290 \\
TutorUp & -0.033 & 0.012 & -2.769 & 0.006 & -0.057 & -0.010 \\
Control & -0.051 & 0.013 & -4.052 & 0.000 & -0.075 & -0.026 \\
DRL Only & -0.039 & 0.014 & -2.768 & 0.006 & -0.067 & -0.011 \\
\hline
\end{tabular}
\end{table}
\FloatBarrier

\begin{table}[H]
\centering
\caption{Pairwise Comparisons (Semantic Similarity)}
\Description{
Pairwise comparisons for semantic similarity. Control vs DRL Only = −0.011 (SE 0.020, z = −0.568, p = 0.570, CI [−0.051, 0.028]). Control vs Full = −0.051 (SE 0.013, z = −4.052, p = 0.000, CI [−0.075, −0.026]). Control vs TutorUp = −0.018 (SE 0.020, z = −0.867, p = 0.386, CI [−0.057, 0.022]). DRL Only vs Full = −0.039 (SE 0.014, z = −2.768, p = 0.006, CI [−0.067, −0.011]). DRL Only vs TutorUp = −0.006 (SE 0.020, z = −0.298, p = 0.766, CI [−0.046, 0.034]). Full vs TutorUp = 0.033 (SE 0.012, z = 2.769, p = 0.006, CI [0.010, 0.057]).
}
\label{tab:lme_pairwise_semantic}
\setlength{\tabcolsep}{2pt}
\begin{tabular}{lcccccc}
\hline
\textbf{Comparison} & \textbf{Coef.} & \textbf{Std.Err.} & \textbf{z} & \textbf{P>|z|} & \textbf{[0.025} & \textbf{0.975]} \\
\hline
Control vs DRL Only & -0.011 & 0.020 & -0.568 & 0.570 & -0.051 & 0.028 \\
Control vs Full & -0.051 & 0.013 & -4.052 & 0.000 & -0.075 & -0.026 \\
Control vs TutorUp & -0.018 & 0.020 & -0.867 & 0.386 & -0.057 & 0.022 \\
DRL Only vs Full & -0.039 & 0.014 & -2.768 & 0.006 & -0.067 & -0.011 \\
DRL Only vs TutorUp & -0.006 & 0.020 & -0.298 & 0.766 & -0.046 & 0.034 \\
Full vs TutorUp & 0.033 & 0.012 & 2.769 & 0.006 & 0.010 & 0.057 \\
\hline
\end{tabular}
\end{table}
\FloatBarrier

\begin{table}[H]
\centering
\caption{Statistical Results (Lexical Similarity)}
\Description{
Statistical results for lexical similarity. Intercept = 0.151 (SE 0.000, z = 340.990, p = 0.000, CI [0.150, 0.152]). TutorUp = −0.016 (SE 0.010, z = −1.527, p = 0.127, CI [−0.036, 0.004]). Control = −0.021 (SE 0.010, z = −2.186, p = 0.029, CI [−0.040, −0.002]). DRL Only = −0.019 (SE 0.010, z = −1.967, p = 0.049, CI [−0.038, −0.000]).
}
\label{tab:lme_overall_lexical}
\begin{tabular}{lcccccc}
\hline
\textbf{Group} & \textbf{Coef.} & \textbf{Std.Err.} & \textbf{z} & \textbf{P>|z|} & \textbf{[0.025} & \textbf{0.975]} \\
\hline
Intercept & 0.151 & 0.000 & 340.990 & 0.000 & 0.150 & 0.152 \\
TutorUp & -0.016 & 0.010 & -1.527 & 0.127 & -0.036 & 0.004 \\
Control & -0.021 & 0.010 & -2.186 & 0.029 & -0.040 & -0.002 \\
DRL Only & -0.019 & 0.010 & -1.967 & 0.049 & -0.038 & -0.000 \\
\hline
\end{tabular}
\end{table}
\FloatBarrier

\begin{table}[H]
\centering
\caption{Pairwise Comparisons (Lexical Similarity)}
\Description{
Pairwise comparisons for lexical similarity. Control vs DRL Only = −0.002 (SE 0.014, z = −0.147, p = 0.883, CI [−0.029, 0.025]). Control vs Full = −0.021 (SE 0.010, z = −2.186, p = 0.029, CI [−0.040, −0.002]). Control vs TutorUp = −0.006 (SE 0.014, z = −0.411, p = 0.681, CI [−0.033, 0.021]). DRL Only vs Full = −0.019 (SE 0.010, z = −1.967, p = 0.049, CI [−0.038, 0.000]). DRL Only vs TutorUp = −0.004 (SE 0.014, z = −0.261, p = 0.794, CI [−0.031, 0.024]). Full vs TutorUp = 0.016 (SE 0.010, z = 1.527, p = 0.127, CI [−0.004, 0.036]).
}
\label{tab:lme_pairwise_lexical}
\setlength{\tabcolsep}{2pt}
\begin{tabular}{lcccccc}
\hline
\textbf{Comparison} & \textbf{Coef.} & \textbf{Std.Err.} & \textbf{z} & \textbf{P>|z|} & \textbf{[0.025} & \textbf{0.975]} \\
\hline
Control vs DRL Only & -0.002 & 0.014 & -0.147 & 0.883 & -0.029 & 0.025 \\
Control vs Full & -0.021 & 0.010 & -2.186 & 0.029 & -0.040 & -0.002 \\
Control vs TutorUp & -0.006 & 0.014 & -0.411 & 0.681 & -0.033 & 0.021 \\
DRL Only vs Full & -0.019 & 0.010 & -1.967 & 0.049 & -0.038 & 0.000 \\
DRL Only vs TutorUp & -0.004 & 0.014 & -0.261 & 0.794 & -0.031 & 0.024 \\
Full vs TutorUp & 0.016 & 0.010 & 1.527 & 0.127 & -0.004 & 0.036 \\
\hline
\end{tabular}
\end{table}
\FloatBarrier

\end{document}